\documentclass[12pt]{article}
\usepackage{amssymb} 
\usepackage{amsfonts}
\newcommand{\beq}{\begin{equation}} \newcommand{\eeq}{\end{equation}}
\begin{document}  
\vskip 2mm

\begin{center}
{\Large\bf Exact analytic two-loop expressions for some QCD observables
in the time-like region}
\vspace{4mm}\\
{\large D.S.~Kourashev}\\

\vspace{0.2cm}
{\it Bogoliubov Lab. Theor. Phys., JINR, Dubna, 141980, Russia;\\
Moscow State Univercity, Physical Department, Moscow, Vorobyevy Gory, 119899, Russia}\\
{\it kourashev@mtu-net.ru}
\end{center}

\abstract{\small
The author of this work have got explicit expressions for the
timelike region at Next-to-Leading-Order (NLO) as for the coupling function
so for QCD observables. These expressions were compared with approximate ones
obtained in terms of ``double logarithms".
Next-Next-to-Leading-Order (NNLO) Pad\'e approximation was also discussed.\par

In this paper some particular role of NLO was emphasised. It is related with
the possibility to express any multi-loop expressions in terms of the two-loop
ones. Thus we can use NLO functions as a basis of some functional expansion,
where higher loop terms incorporation affects the expansion coefficients only.

}

\section{Introduction}
Renorm-group equation (RGE) for QCD coupling function $\bar \alpha _s$ could
be written as 
\begin{eqnarray}
x\frac{\partial \bar \alpha _s}{\partial x}=-(\beta_0 \bar \alpha _s^2+
\beta_1 \bar \alpha _s^3+\beta_2 \bar \alpha _s^4+...)
\end{eqnarray}
with $\beta _n$ corresponding to $(n+1)$-loop contribution. Its solution 
has a very simple form in the one-loop case: $\bar \alpha _s^{(1)}(x)=
\frac 1{\beta_0 \ln x}$.
At NLO (two-loop approximation) it can be expressed through the Lambert
function~\cite{magradze, magradze1, grunberg, lambert}
\begin{eqnarray}\label{a2}
\bar \alpha^{(2)}_s(x)=-\frac {\beta_0}{\beta_1}\frac 1{1+W(z)}\ \ \mbox{, } \ \ \
z \equiv -\frac {\beta_0^2}{\beta_1 e} x^{ -\frac{\beta_0^2}{\beta_1} }.
\end{eqnarray}
defined by a transcendental equation $W(z)e^{W(z)}=z$. Here and further $x=%
\frac{Q^2}{\Lambda ^2}$, $\Lambda $ is the scale parameter defining unphysical
singularities positions.
\par

It should be mentioned that all RGE solutions obtained perturbatively
have unphysical singularities
such as ``ghost poles". In the two- and three-loop cases the cut
lying in infrared
region emerges. Hence, the coupling behaviour in this region can't be described
this way. Analytic Approach suggested four years ago by D.V.~Shirkov and
I.L.~Solovtsov~\cite{dv96,dv97,dv98} allowed to obtain analytic
expressions for the running coupling with the same ultraviolet
asymptotic that ordinary RGE solutions have but with a stable infrared
behaviour. Afterwards, this procedure was slightly corrected and generalised
by D.V.~Shirkov~\cite{dv00, dv00b}. In the Sec.2 some basic facts from this
approach were stated and also some two-loop calculations results were
presented.\par

In the Sec.3 the Analytic Approach was applied for the two- and three- loop cases.
It was proved in this work that we can obtain exact expressions for the
two-loop case and Pad\'e approximated third order $\beta$-function.
These results were compared with the ones obtained using "double logarithms"
approximation.\par

As follows from the publication~\cite{moi} it's possible to express the coupling
constant for any-loop case as a power series of the two-loop coupling. It's
also possible to apply Analyticization Procedure to this expansion.
The opportunity to express some arbitrary coupling function or observable
in terms of two-loop functions is discussed in the Sec.4.

\section{Modified Analyticization Procedure}
                                  
Standard Analytic Approach allows us to get rid of unphysical singularities and
make it possible to research the coupling (and observables also) behaviour
in the low energy region. Moreover it reduces the scheme dependence
considerably. However it is also necessary to incorporate quark thresholds
and to construct the procedure allowing
to transfer the coupling (and observables also) to the time-like
region. These problems were solved by D.V.~Shirkov~\cite{dv00, dv00b} in
his recent works.

        \subsection{$Q^2$-channel}
As it had been proposed earlier by the Analyticization procedure, it is possible
to make the coupling analytic in a whole $Q^2$ complex plane except
the cut on the real negative axis. However, this approach does not take into
account the
thresholds problem. Analyticization procedure is based on K\"allen-Lehmann
spectral representation, it leads to the modified coupling function that
is devoid of unphysical singularities. Besides, it has the correct asymptotic
behaviour consistent with the perturbative result. The spectral representation
has the form
\begin{eqnarray}
\left\{ a(x)\right\} _{an}=\frac 1\pi \int\limits_0^\infty
\frac{\rho(\sigma)}{\sigma +x}d\sigma,
\end{eqnarray}
the spectral density can be taken as
$\rho(\sigma)=\Im \bar\alpha_s^{f=3}(-\sigma )$ that
is natural for the low energy processes analysis when there are three
active quarks ($f=3$). Here, the new coupling function is defined through an
imaginary part of usual, RG invariant, effective coupling
$\bar{\alpha_s}$ continued on the physical cut.  \par

However, it is also possible to involve the threshold matching
procedure. It is well known that one can relate the couplings
for the regions with a different flavours numbers. One should
match $\bar\alpha_s^{f=n-1}(M_n^2)$ and $\bar\alpha_s^{f=n}(M_n^2)$ by
the proper scale parameter redefinition. As a result,
spectral density is discontinuous~\cite{dv00} and can be presented as
\begin{eqnarray}
\rho(\sigma)=\rho^{f=3}(\sigma)+\sum_{f>3}\theta(\sigma-M_f^2)(\rho^f(\sigma)-
\rho^{f=3}(\sigma)).
\end{eqnarray}
Of course, the coupling obtained from this density is continuous.

        \subsection{$s$-channel}
Some interesting experiments data correspond to the timelike region (s-channel)
that leads to the necessity of the relevant redefinition of coupling function.
Some approximate estimates, such as
${\mathfrak A}(s)=\bar{\alpha_s}(s)$ or
${\mathfrak A}(s)=|\bar{\alpha_s}(-s)|$,
were used earlier for this purpose. These estimates are suitable for
qualitative analysis only in the region higher than $5GeV$,
and they can not reflect the correct analytical
properties. The difference between these estimates and exact
expressions is within several per cents in this region and this precision
can be accurate enough for some calculations.\par

However, it is better to use the ``dipole representation" for the Adler
function in terms of an observable $R(s)$ in the timelike region
\begin{eqnarray}
D(Q^2)=Q^2\int_0^{\infty} \frac{ds}{(s+Q^2)^2}R(s)
\end{eqnarray}
and an inverse expression
\begin{eqnarray}
R(s)=\frac{i}{2\pi}\int_{-s-\i\epsilon}^{-s+\i\epsilon}\frac{D(x)}x dx.
\end{eqnarray}
This transformation leads to the coupling function definition for s-channel.
It can be implemented as
\begin{eqnarray}\label{A_s}
{\mathfrak A}=\frac{i}{2\pi}\int_{-s-\i\epsilon}^{-s+\i\epsilon}\frac{\alpha_s(x)}x dx.
\end{eqnarray}
It can also be expressed in terms of the spectral density~\cite{milsol}
\begin{eqnarray}
{\mathfrak A}(s)=\frac{1}{\pi}\int_{s}^{\infty}\frac{d\sigma}\sigma\rho(\sigma).
\end{eqnarray}\par
These formulae have been applied to one-loop case by K.~Milton and
O.~Solovtsova~\cite{milsola}.
Corresponding expression for ${\cal A}(s)$ is
\begin{eqnarray}
{\mathfrak A}=\frac1{2\beta_0}-\frac1{\pi\beta_0}\arctan \frac{\ln \frac{s}{\Lambda^2}}{\pi}
\end{eqnarray}
Higher orders was also obtained in this paper.

        \subsection{Thresholds}
As follows from the work by D.V.~Shirkov~\cite{dv00}, the effective
s-channel coupling ${\mathfrak A}(s)$
differs from ${\mathfrak A}^f$ by a {\it shift constant}
\begin{eqnarray}
{\mathfrak A}(s)={\mathfrak A}^f+c^f \mbox{ at } M_f^2\le s\le M_{f+1}^2,
\end{eqnarray}
the {\it shift constants} can be easily calculated
\begin{eqnarray}
c^{f-1}=c^f+{\mathfrak A}^f(M_f^2)-{\mathfrak A}^{f-1}(M_f^2).
\end{eqnarray}
These constants values correspond to 4th, 5th and 6th quarks
contributions. The shift constants provide the continuity of the analytic
coupling function. Indeed, in spite of discontinuous spectral density we
should get the continuous expressions from the spectral representation.\par

Here are some typical values calculated at NLO:
$$\begin{array}{ccccccc}\label{c_i}
\Lambda_3, MeV & \Lambda_4, MeV & \Lambda_5, MeV & \Lambda_6, MeV & c^3 & c^4 & c^5\\
300 & 251 & 174 & 72 & 0.0107 & 0.0028 & 0.00003\\
400 & 346 & 249 & 108 & 0.0149 & 0.0035 & 0.00003\\
\end{array}$$
Note that these contributions gather about 5 per cent 
to the effective coupling in the low energy region and can considerably
influence the scale parameter value.  \par

Quite analogously, {\it second shift constant} and
{\it third shift constant} are defined
\begin{eqnarray}
{\mathfrak A}_i(s)={\mathfrak A}_i^f+c_i^f \mbox{ at } M_f^2\le s\le M_{f+1}^2,
\mbox{ where}
\end{eqnarray}
\begin{eqnarray}
c_i^{f-1}=c_i^f+{\mathfrak A}_i^f(M_f^2)-{\mathfrak A}_i^{f-1}(M_f^2), \mbox{ i=2,3}.
\end{eqnarray}
Some typical values of $c_2$ and $c_3$ are presented below:
$$\begin{array}{cccccccc}\label{c23_i}
\Lambda_4, MeV & c_2^3 & c_2^4 & c_2^5 & c_3^3 & c_3^4 & c_3^5\\
300 & -0.0021 & -0.0024 & -0.0008 & 0.0026 & 0.0005 & 0.00002\\
500 & 0.0007 & -0.0024 & -0.0004 & 0.0047 & 0.00009 & 0.00003\\
\end{array}$$

\section{Two-loop and three-loop applications}

        \subsection{Exact NLO results}

At NLO we can even obtain the analytical expressions
for the s-channel Analyticized coupling function. As it was mentioned the
two-loop renormgroup equation solution can be expressed in terms of the
Lambert function (\ref{a2}).
This representation also allowed us to obtain
${\mathfrak A}_i(s)$ functions {\it exactly}
\begin{eqnarray}
{\mathfrak A}(s)=-\frac {\beta_0}{\beta_1} + \frac 1{\pi}\Im \frac {1}{\beta_1 \bar
\alpha_s(-s)}.
\end{eqnarray}

\begin{eqnarray}
{\mathfrak A}_2(s)=\frac 1{\pi \beta_1} \Im \ln (1+\frac{\beta_1}{\beta_0}
\bar \alpha_s(-s)).
\end{eqnarray}

\begin{eqnarray}
{\mathfrak A}_3(s)=- \frac{\beta_0}{\beta_1} \frac 1{\pi \beta_1} \Im
\left\{ \ln (1+ \frac{\beta_1}{\beta_0} \bar \alpha_s(-s)) -
\frac{\beta_1}{\beta_0} \bar \alpha_s(-s)\right\}.
\end{eqnarray}

\begin{eqnarray}
{\mathfrak A}_4(s)=\left(- \frac{\beta_0}{\beta_1}\right)^2
\frac 1{\pi \beta_1} \Im
\left\{ \ln (1+ \frac{\beta_1}{\beta_0} \bar \alpha_s(-s)) -
\frac{\beta_1}{\beta_0} \bar \alpha_s(-s)+
\frac{\beta_1^2}{2\beta_0^2} \bar \alpha_s^2(-s)\right\}, etc.
\end{eqnarray}\par

These expressions are obtained by using the contour integral~(\ref{A_s}) in
the generalized form
\begin{eqnarray}
{\mathfrak A}_n=\int_{s-\i\epsilon}^{s+\i\epsilon}\frac{\alpha_s^n(x)}x dx.
\end{eqnarray}\par
The integration can be implemented with rather simple variables substitutes.
It's interesting that ${\mathfrak A}_n(s)$ can be presented as $n-2$ residual terms
of the Taylor expansion of
$\frac 1{\pi \beta_1} \Im \ln (1+\frac{\beta_1}{\beta_0} \bar \alpha_s(-s))$
over the
powers of $\bar \alpha_s(-s)$ multiplied by $- \frac{\beta_0}{\beta_1}$.

         \subsection{Three-loop Pad\'e approximation}
The renorm-group equation at NNLO can not be solved explicitly. However
the Pad\'e-approximated $\beta $-function
\begin{eqnarray}\label{bpade}\beta_{Pad\acute e}=-\beta_0
\bar \alpha_s^2\left(1+\frac{\beta_1 \bar \alpha_s}
{\beta_0-\frac{\beta_0 \beta_2}{\beta_1}a}\right)=\beta_0 \bar \alpha_s^2+
\beta_1 \bar \alpha_s^3+\beta_2 \bar \alpha_s^4+\frac{\beta_0 \beta_2^2}{\beta_1}
\bar \alpha_s^5\,.\,.\,.\end{eqnarray}
can be used in this
case and the solution can be also obtained through Lambert function~\cite{pade}
\begin{eqnarray}\label{pade}
a_{Pad\acute e}^{(3)}(x)=-\frac {\beta_0}{\beta_1}\frac 1{1-c+W(z)}
\mbox{, where }\ \
z=-\frac 1{eb} e^c x^{-\frac 1b},\,
b=\frac{\beta_1}{\beta_0^2}, c=\frac{\beta_0\beta_2}{\beta_1^2} .
\end{eqnarray}

Without Pad\'e%
-approximation even in the three-loop case we can't get analytically the RGE
solution for both coupling and observables. So it is natural to use the
the expansion through two-loop
functions to analyse multi-loop functions.\par

One can obtain the s-channel coupling in this
case exactly like in the two-loop case
\begin{eqnarray}
{\mathfrak A}(s)=-\frac 1{\pi\beta_0} \Im\left\{ \frac{\ln\left( -
\frac 1c (1+W(z))+1\right)}{1-\frac 1c } +
\frac{\frac 1c \ln(-W(z))}{1-\frac 1c }
\right\}
\ \mbox{, }\ \
z=-\frac 1{eb} e^c (-s)^{-\frac 1b}.
\end{eqnarray}\par

         \subsection{Using power expansion over the two-loop
                functions for NNLO analysis}
We can also get the three-loop functions without Pad\'e approximation
using the method suggested in the work~\cite{moi}. The expansion~(\ref{nlo_exp})
allows us to obtain the following expressions
\begin{eqnarray}
{\mathfrak A}^{NNLO}(s)=k_1 {\mathfrak A}^{(2)}(s) + k_2 {\mathfrak A}_2^{(2)}(s) +
k_3 {\mathfrak A}_3^{(2)}(s) + k_4 {\mathfrak A}_4^{(2)}(s) + k_5 {\mathfrak A}_5^{(2)}(s) +...,
\end{eqnarray}\par
for NNLO case $k_1=1, k_2=0, k_3= \frac{\beta_2}{\beta_0^3}, k_4=0,
k_5= \frac{5\beta^2_2}{3\beta_0^6}$. This case differs from the
case with Pad\'e-transformed $\beta$-function as the latter originates some
``trail" of higher order terms. However this difference is quite small (when
comparing new shift constants the difference between NLO and NNLO is
about 15 per cents, while it is only 1 per cent between NNLO and ``Pad\'e"
case).

        \subsection{Comparing exact two-loop functions with approximate
        ones}
Two-loop RGE solution (\ref{a2}) is explicit and as it is argued in the
present work the Lambert function is also very convenient for the timelike
region research. In spite of these facts as a rule one takes some approximate
expressions like
\begin{eqnarray}\label{a_app}
\alpha_{s app}^{(2)}(x)=\frac 1{\beta_0 \ln(x)}\left\{\left(1-
\frac {\beta_1}{\beta_0^2} \frac{\ln(\ln(x))}{\ln(x)}
+\frac{\beta_1^2}{\beta_0^4 \ln^2(x)}\left((\ln(\ln(x))-\frac 12\right)^2-\frac 54
\right)\right\}.
\end{eqnarray}
As it was proved we can express this function in terms of (\ref{a2})
\begin{eqnarray}\label{a_exp}
\alpha_{s app}^{(2)}(x)=\alpha^{(2)}_s(x)+k(x)
\left(\alpha^{(2)}_s(x)\right)^4+O\left(\alpha^{5}_s(x)\right)
\end{eqnarray}
Here $k(x)$ is some limited function. Some typical values of this function
are\footnote{We've taken $\Lambda_3=300MeV$}:
$$\begin{array}{cc}\label{kx}
Q, MeV & k\left(\frac{Q^2}{\Lambda^2}\right) \\
600 & -0.7 \\
900 & -0.78 \\
1200 & -0.83 \\
1500 & -0.85 \\
2100 & -0.91 \\
\end{array}$$
This terms proportional to the fourth power of coupling function contributes
for about 4 per cent difference in 1GeV region. We have to note that this
can not be corrected by the scale parameter choice as it would originate
nonzero coefficients before 2nd and 3rd powers of $\alpha^{(2)}_s$ in the
expansion~(\ref{a_exp}).

\section{Obtaining the multi-loop functions in terms of two-loop ones}

        \subsection{Expansion}
Suppose some observable can be expressed as a power series of the coupling
constant
\begin{eqnarray}
F(Q^2)=F_1 \bar \alpha _s(Q^2) + F_2 \bar \alpha _s^2(Q^2) +
F_3 \bar \alpha _s^3(Q^2)...
\end{eqnarray}
However we face unphysical singularities, wrong analytic behaviour in the
infrared (IR) region. We can not consider early attempts to express
this observable in s-channel like
\begin{eqnarray}
F_s(s)=F'_1 \bar \alpha _s(s) + F'_2 \bar \alpha _s^2(s) +
F'_3 \bar \alpha _s^3(s)...
\end{eqnarray}
to be very successful. Indeed, it involves the so called $\pi^2$-terms and
leads to the expansion coefficients
augmenting~\cite{radyushkin, pivovarov, kataev}, this way also preserves unphysical
singularities.
As a result of applying the Analyticization procedure 
the analytic coupling function and the functional
expansion of observables both in $Q^2$- and $s$- channels can be obtained
\begin{eqnarray}
F(Q^2)=F_1 {\cal A}(Q^2) + F_2 {\cal A}_2(Q^2) + F_3 {\cal A}_3(Q^2)...
\end{eqnarray}
\begin{eqnarray}
F_s(s)=F_1 {\mathfrak A}(s) + F_2 {\mathfrak A}_2(s) + F_3 {\mathfrak A}_3(s)...
\end{eqnarray}
After all, the Analytic approach leads to the scheme dependence reduction that is
very important physical consequence.

        \subsection{The expansion over two-loop functions}
As follows from the paper~\cite{moi} we can obtain the expressions for
multi-loop functions in terms of two-loop functions. Without analyticization
we have the an expansion over the coupling function powers like
\begin{eqnarray}
a^{multi-loop}=a^{(2)} + k_2 a^{(2) 2}  + k_3 a^{(2) 3} + ...
\end{eqnarray}
Here we can easily apply the analytic approach. It leads to some modifications
when we've got the functional expansion instead of the power one
\begin{eqnarray}
A_{multi-loop}=A^{(2)} + k_2 A_2^{(2)}  + k_3 A_3^{(2)} + ...
\end{eqnarray}
Thus two-loop functions can be considered as a minimal basis of any orders
perturbation expansions. So any observable $F$ (devoid of anomalous dimensions)
can be presented over these functions in arbitrary loop order
\begin{eqnarray}\label{nlo_exp}
F=F_1 A^{(2)} + F_2 A_2^{(2)}  + F_3 A_3^{(2)} + ...
\end{eqnarray}
Some similar result (observables expansion as a power expansion over two-loop
functions) was obtained independently by C.~Maxwell~\cite{maxwell}.

\section{Conclusions}

In this paper the author has got an 
{\it exact} expressions for analytic running coupling at the two-loop order.
These expressions differ from approximate ones by a term proportional to
the fourth power of the coupling function. Timelike functions were also
obtained just in the terms of Lambert function.
As it was shown higher orders loop approximations can be obtained
as an expansion over two-loop functions.
We need to be only provided with $\beta$-coefficients.\par

This approach allows us to get obtain accurate theoretical predictions of 
the coupling function consistent with its analyticity properties.
As it was argued the difference between exact expressions and approximate ones
in the infrared region is high enough to be considered.
\par

\bigskip %

{\large\bf Acknowledgements}
\vspace{0.1cm}

It should be mentioned that significant part of this paper results
(exact NLO expressions) were
obtained independently almost at the same time by B.~Magradze
(see~\cite{magradze2}). The author considering this paper as the first
step hopes to continue this research jointly with Badri Magradze.
The author would like to thank D.V.~Shirkov for his
invaluable help in the creation of this paper.
This work was partially supported by Russian Foundation for Basic Research
grant No 99-01-00091.


\begin{thebibliography}{99}

\bibitem{magradze} B.A.~Magradze ``The Gluon Propagator in Analytic Perturbation
Theory", Talk given at 10th Intern. Seminar on High-Energy Physics
(Quarks 98), Suzdal, Russia, 18-24 May 1998, Preprint G-TMI-98-08-01,
Tbilisi, 1998, hep-ph/9808247;

\bibitem{magradze1} B.A.~Magradze, Int.J. Mod. Phys.A 15 (2000) 2715-2733,
hep-ph/9911456;

\bibitem{grunberg}  E.~Gardi, G.~Grunberg, M.~Karliner ``Can the QCD running
coupling have a causal analyticity structure?'', hep-ph/9806462;


\bibitem{lambert} R.M.~Carless, G.H.~Gonnet, D.E.G.~Hare, D.J. Jeffrey and
D.E.~Knuth, ``On the Lambert W function", Advances in Computational
Mathematics, 5 (1996) 329;

\bibitem{dv96} D.V.~Shirkov, I.L.~Solovtsov, JINR Rapid Comm.,
No. 2[76]-96,
hep-ph/9604363;

\bibitem{dv97} D.V.~Shirkov, I.L.~Solovtsov ``Analytic Model for the QCD
Running Coupling with Universal $\bar\alpha _S(0)$  Value,
Phys. Rev. Lett. 79 (1997) 1209-12, hep-ph/9704333;

\bibitem{dv98}  D.V.~Shirkov ``Renormgroup, Causality, Non-power
Perturbation Expansion in QFT'', Teor.Mat.Fiz 119 (1999) 55, hep-th/9810246;

\bibitem{dv00} D.V.~Shirkov ``Towards the corellated analysis of observables
in perturbative QCD", JINR preprint E2-2000-46; hep-ph/003242;

\bibitem{dv00b} D.V.~Shirkov ``The $pi^2-$terms in the $s$-channel QCD
observables"; hep-ph/0009106;

\bibitem{milsol} K.A.~Milton and I.L.~Solovtsov, Phys. Rev. D 55 (1997)
5295-5298;

\bibitem{milsola} K.A.~Milton and O.P.~Solovtsova, Phys. Rev. D 57 (1998)
5402-5409;

\bibitem{radyushkin} A.Radyushkin, Dubna JINR preprint E2-82-159 (1982);
see also JINR Rapid Comm. No. 4[78]-96 (1996) pp9-15 and hep-ph/9907228.

\bibitem{pivovarov} N.V.~Krasnikov and A.A.~Pivovarov, Phys. Lett. 116 B (1982)
168-170;

\bibitem{kataev} A.L.~Kataev, V.V.~Starshenko, Mod.Phys.Lett. {\bf A10} (1995) 235;

\bibitem{moi} D.S.~Kourashev, hep-ph/9912410;

\bibitem{maxwell} C.J.~Maxwell ``Complete Renormalization Group Improvement -
Avoiding Scale Dependence in QCD Prediction", hep-ph/9908463;

\bibitem{pade} J.~Ellis, E.~Gardi, M.~Karliner, M.A.~Samuel ``Pade
Approximants, Borel Transforms and Renormalons: the Bjorken Sum Rule as a
Case Study", Phys. Lett. {\bf B366} (1996) 268, hep-ph/9509312;

\bibitem{magradze2} B.~Magradze ``The QCD coupling up to third order in standard
and analytic perturbation theories", JINR preprint E2-2000-222, hep-ph/0010070.


\end{thebibliography}
\end{document}